# Echoes from Ancient Supernovae in the Large Magellanic Cloud

Armin Rest[1], Nicholas B. Suntzeff[1], Knut Olsen[1], Jose Luis Prieto[2], R. Chris Smith[1], Douglas L. Welch[3], Andrew Becker[4], Marcel Bergmann[5], Alejandro Clocchiatti[6], Kem Cook[7], Arti Garg[8], Mark Huber[7], Gajus Miknaitis[4], Dante Minniti[6], Sergei Nikolaev[7], & Christopher Stubbs[8]

[1]*Cerro Tololo Inter-American Observatory, National Optical Astronomy Observatory[9], La Serena, Chile*

[2]*Dept. Astronomy, Ohio State University, Columbus, OH 43210 USA*

[3]*Dept. Physics & Astronomy, McMaster University, Hamilton, ON, L8S 4M1, Canada*

[4]*Dept. Astronomy, University of Washington, Seattle 98195 USA*

[5]*Gemini Observatory[9], La Serena, Chile*

[6]*Dept. Astronomia y Astrofisica, Pontifica Universidad Católica de Chile, Santiago, Chile*

[7]*Lawrence Livermore National Laboratory, Livermore, CA 94550 USA*

[8]*Dept. of Physics and Harvard/Smithsonian Center for Astrophysics, Harvard University, Cambridge, MA 02138 USA*

[9] *Based on observations obtained at NOAO, operated by the Association of Universities for Research in Astronomy, Inc. (AURA) under cooperative agreement with the NSF.*

**In principle, the light from historical supernovae could still be visible as scattered-light echoes even centuries later[1-6]. However, while echoes have been discovered around some nearby extragalactic supernovae well after the explosion[7-13], targeted searches have not recovered any echoes in the regions of historical Galactic supernovae[14-16]. The discovery of echoes can allow us to pinpoint the supernova event both in position and age and, most importantly, allow us to acquire spectra of the echo light to type the supernova centuries after the direct light from the explosion first reached the Earth. Here we report on the discovery of three faint new variable surface brightness complexes with high apparent proper motion pointing back to well-defined positions in the Large Magellanic Cloud (LMC). These positions correspond to three of the six smallest (and likely youngest) previously catalogued supernova remnants, and are believed to be due to thermonuclear (Type Ia) supernovae[17]. Using the distance and proper motions of these echo arcs, we estimate ages of 610 and 410 yr for the echoes #2 and #3.**

As part of the SuperMACHO microlensing survey, we have been monitoring the central portion of the LMC every other night for three months each year over the last four years (2001-4). Using an automated pipeline, we subtract point-spread-function matched template images from the recent epoch images. The resulting difference images are remarkably clean of the constant stellar background and are ideal for searching for variable objects.

The well-known echo of SN1987A shown in Figure 1 was trivial to recover in the difference images with our pipeline. The high apparent motion of the echoes, often superluminal, allows simple detection in difference images. To search for very faint echoes, we have examined by eye all the variable objects discovered by our automatic pipeline. We found a number of very faint linear structures that had high proper motions with vector directions inconsistent with the 1987A echo. For each structure, we estimated a vector direction as shown in Figure 2. Figure 3 shows the echo vectors extrapolated backward in time pointing to three well-defined positions as the origins of the echo complexes. The origins of the four echo complexes are listed in Table 1. The three unidentified echo origins correspond within arcminutes of the positions of known supernova remnants (SNR)[18] and also correspond to three of the six youngest SNRs[17]. These three SNRs are precisely the three that are classified as likely Type Ia events based on the X-ray emission spectra.

Given the positional match with young SNRs and the high apparent proper motions of the variable diffuse light, we conclude that these newly detected structures are likely to be scattered light echoes from Type Ia supernovae in the LMC. Planned spectroscopy of the brightest knots in the three echo complexes should allow us to determine the type of the supernovae and confirm the classifications from the X-ray studies.

The theory of supernova light echoes (whereby we mean the actual scattered light echo rather than fluorescence or dust re-radiation) predicts that light echoes can be seen even centuries after the first arrival of light from the explosion. Using a differential form of equation 7 for surface brightness[19], we find for two different supernovae:

$$\Sigma_2 = \Sigma_1 + (V_{2\_SN} - V_{1\_SN}) - 2.5\log_{10}(r_1 t_1 / (r_2 t_2)) - 2.5\log_{10}(\Phi_2/\Phi_1)$$

where $\Sigma$ is the echo surface brightness, $V_{\_SN}$ is the supernova magnitude at maximum, $r$ is the echo to supernova distance, $t$ is the time between explosion and echo observation, and $\Phi$ is the Henyey-Greenstein phase function. Here we assume that the SN light pulse duration is the same for the two supernovae, and that the composition, density, and thickness of the dust sheets producing the echoes are identical. We also calculate the $\Phi$ function with forward scattering (g=0.6), and only include the angular terms. Scaling from the brightest echo knot of SN1987A at 19.3 mag arcsec$^{-2}$, we find that a 500 year old Type Ia SN that exploded 150pc behind a face-on dust sheet would produce a light echo with a surface brightness of 22.5 mag arcsec$^2$ at an angular distance of 0.29º (250pc radial distance from the SN) assuming a Type Ia supernova was 3.5mag brighter than SN1987A. At 1000 years, the echo would be 24 mag arcsec$^2$ at an angular distance of 0.5º or 420pc from the explosion site. These surface brightness estimates are consistent with the echoes discovered here.

Supernova light echoes can be used to measure the structure and nature of the interstellar medium[4, 20, 21] and, in principle, can be used to measure geometric distances[22]. The geometric relationship is $\rho = (ct(2z + ct))^{1/2}$ where $\rho$ is the apparent projected radius of the light echo on the sky, $z$ is the distance from the supernova to the dust sheet, and $t$ is the time since peak brightness of the source. Given the known distance to the LMC and time of explosion, the echoes in Figure 1 can be used to map out the structure of the dust[23].

What are the ages of the supernovae producing these echoes? A Type Ia SN in the LMC would reach an apparent magnitude of $V \sim -0.5$ and would be the second or third brightest star in the southern sky for a few weeks. Lower limits on the supernova ages can be set from the absence of reported bright supernovae since the establishment of the Royal Observatory at the Cape in 1820. An independent lower limit of 300 yr can be derived from the sizes of these SN remnants assuming an unrealistic large constant shock velocity of 10,000 km s$^{-1}$.

We can use the apparent expansion velocity to crudely measure the ages of the supernova echoes. A simple differentiation of the formula above gives $v=c(z+ct)/\rho$ where $v$ is the expansion velocity assuming the dust plane is perpendicular to the line of sight and $c$ is the speed of light. Solving the two equations simultaneously, we find the age for echo 2 is 600 ± 200 yr with the dust 570 ± 180pc in front of the SN based on 9 arcs, and for echo 3, an age of 400 ± 120 yr with the dust 340 ± 160pc in front of the SN based on 6 arcs. Echo 4 only had one arc with a superluminal velocity, giving an age of 860 yr. The alternative solutions to the equations gave ages greater than 2500 yr, which are excluded based on upper limits of less than 1000 to 1500 yr from the optical and X-ray observations[24]. As a check on this technique, we measured an age for the SN1987A echo of 15.9 ± 1.4 yr from 39 echo arc positions, which is consistent with the age of 1987A at the epoch of observation of 14.8 yr.

The uncertainties quoted above are the standard deviation of estimates from the different arcs. The uncertainties in the proper motions, which are typically 0.1 arcsec yr$^{-1}$, propagate to age uncertainties of less than 50 yr. The largest uncertainty in the age estimates comes from the unknown inclinations of the dust sheets (assumed to be zero, or perpendicular to the line of sight). Allowing for inclinations leaves the upper limit on the ages unbounded, but lower limits can still be derived. If the dust sheets have inclinations of less than 60 degrees, we find lower limits of 400 yr, 250 yr, ad 380 yr for the ages of echo 2, 3, and 4 respectively.

Also intriguing is the opportunity they provide for directly observing the spectral light from the historical supernovae themselves as Zwicky[25] suggested in 1940. Precise image subtraction techniques on nearby galaxies and in our own Galaxy with modern digital images can reach much fainter surface brightness limits than the early photographic surveys and allow us to find echoes from supernovae as old as 1000 years or more. With the discovery of a bright echo knot, we might be able today take a spectrum, representing the time average of the light at maximum, of the Tycho, Kepler, SN1006, or Cas A supernova. As an example, for a dust sheet 400pc in front of the Tycho SN with $V_{max}$=-6.5, a distance of 3.1kpc, and knots of densities similar to the highest density sheets near SN1987A, the surface brightness would be 22 mag arcsec$^{-2}$.

The arc would be at 6.5º from the Tycho SNR and would move at 30" yr$^{-1}$. Scaling the typical echo width from the LMC, the Galactic echo would be ~30" wide. A survey utilizing digital subtraction over an area of 100 sq-degree could be able to recover these moving arcs.

Style tag for received and accepted dates (omit if these are unknown).

**Acknowledgements** C.S. thanks the National Science Foundation, the McDonnell Foundation, and Harvard University for their support of the SuperMACHO project. D.W. acknowledges support from the Natural Sciences and Engineering Research Council of Canada (NSERC). The work of K.C., M.H. and S.N. was performed under the auspices of the U.S. Department of Energy, National Nuclear Security Administration by the University of California, Lawrence Livermore National. A.C. acknowledges support from FONDECYT. DM was partially supported by FONDAP. J.P. was funded by the OSU Astronomy Department Fellowship.

**Compelling interests statement** The authors declare that they have no compelling financial interests.

**Correspondence** and requests for materials should be addressed to N.S. (e-mail: nsuntzeff@ctio.noao.edu).).

**Figure 1.** The light echoes from SN 1987A. The data, taken at the CTIO 4m Blanco telescope with the MOSAIC imager in the *VR* filter, were used to make this difference image with epoch 2004.97 minus 2001.95 data, representing 17.8 and 14.8 years after the explosion. Our SuperMACHO survey covers 24 sq-degrees in 68 pointings in an approximate rectangle 3.7° by 6.6° aligned with the LMC bar. The images are taken through our custom "*VR*" filter ($\lambda_c$=625nm,

Δλ=220nm) with exposure times of 60s to 200s, depending on the stellar densities. The field is 13.8' by 18.4' with N up and E left. White represents flux enhancements in the 2004 image and black in the 2001 image. Faint echo arcs can be seen as far out as 6.6' and 7.3' from the explosion site, or 0.9 and 1.1kpc in front of SN 1987A. The *VR* surface brightness varies from 19.8 to a limit of ~24 mag arcsec$^{-2}$ with one knot as bright as 19.3 mag arcsec$^{-2}$. The widths of the echoes are resolved, and typically ~2.5" across.

**Figure 2.** Arcs of light echoes in the Large Magellanic Clouds from previously unseen supernovae. Panel 1 (upper left) shows the unsubtracted (template) image which includes the cluster Hodge 243. Panel 2 (upper right) shows how cleanly the field subtracts with data taken 50d earlier. The next three panels show the echo motion 1, 2, and 3 years after the template date. White represents positive flux in the present epoch image and black in the template image. The vector motions are plotted in Panel 6 (lower right). Each echo is fit with a straight line (red). The apparent proper motion is given by the yellow vector and extrapolated backwards (blue). The size of the yellow vector is proportional to the length of the echo segment fit. Saturated stars are masked out with grey circles. A number of faint variable stars appear as black or white spots. The vector was defined to be perpendicular to a linear fit to an echo segment, with the direction given by the proper motion. Typical proper motions range from 0.5-2.4" yr$^{-1}$ which, at the angular scale of the LMC of 0.77 light-year arcsec$^{-1}$ makes many of these structures have apparent superluminal velocities. The surface brightness ranges from 22.3 mag arcsec$^{-2}$ down to our limit of detection at 24 mag arcsec$^{-2}$. These echoes are located in echo complex #2, at RA, Dec=(05:16:06,-69:17:07, J2000). Each panel is 80" on a side with N up and E to the left.

**Figure 3.** A plot of the light echo vectors in the LMC. The vectors have the same meaning as in Figure 2. The centres of the echo complexes are indicated by yellow circles. The lengths of the yellow vectors are 100x the length of the echo arc. The source on the left marked with a star is SN1987A. The green circles are the location of historical novae, and the red circles are the supernova remnant locations[25]. Evidently, the three unknown echo complexes point to three catalogued supernova remnants. We have estimated the position of the crossing point of the vectors by calculating the crossings of all pairs of vectors in each group excluding any echo pair with a separation of less than 10".

**Table 1: Positions of Supernova Echo Origins in the LMC**

| Echo complex | RA | dec | position error | δr | SNR name |
|---|---|---|---|---|---|
| 1 | 05:35:30 | -69:16 | 0.1 | 0.2 | SN1987A |
| 2 | 05:19:14 | -69:04 | 1 | 2.5 | 0519-69.0 |
| 3 | 05:11:17 | -67:31 | 1 | 10.0 | 0509-67.5 |
| 4 | 05:09:19 | -68:42 | 2 | 2.3 | 0509-68.7 (N103B) |

Position errors, based on the intersection of the echo vectors, are given in arcminutes. δr, the distance between the tabulated echo origin and SNR, is given in arcminutes. Coordinates are equinox J2000. The error in the centroid was estimated from the averaged vector crossings.

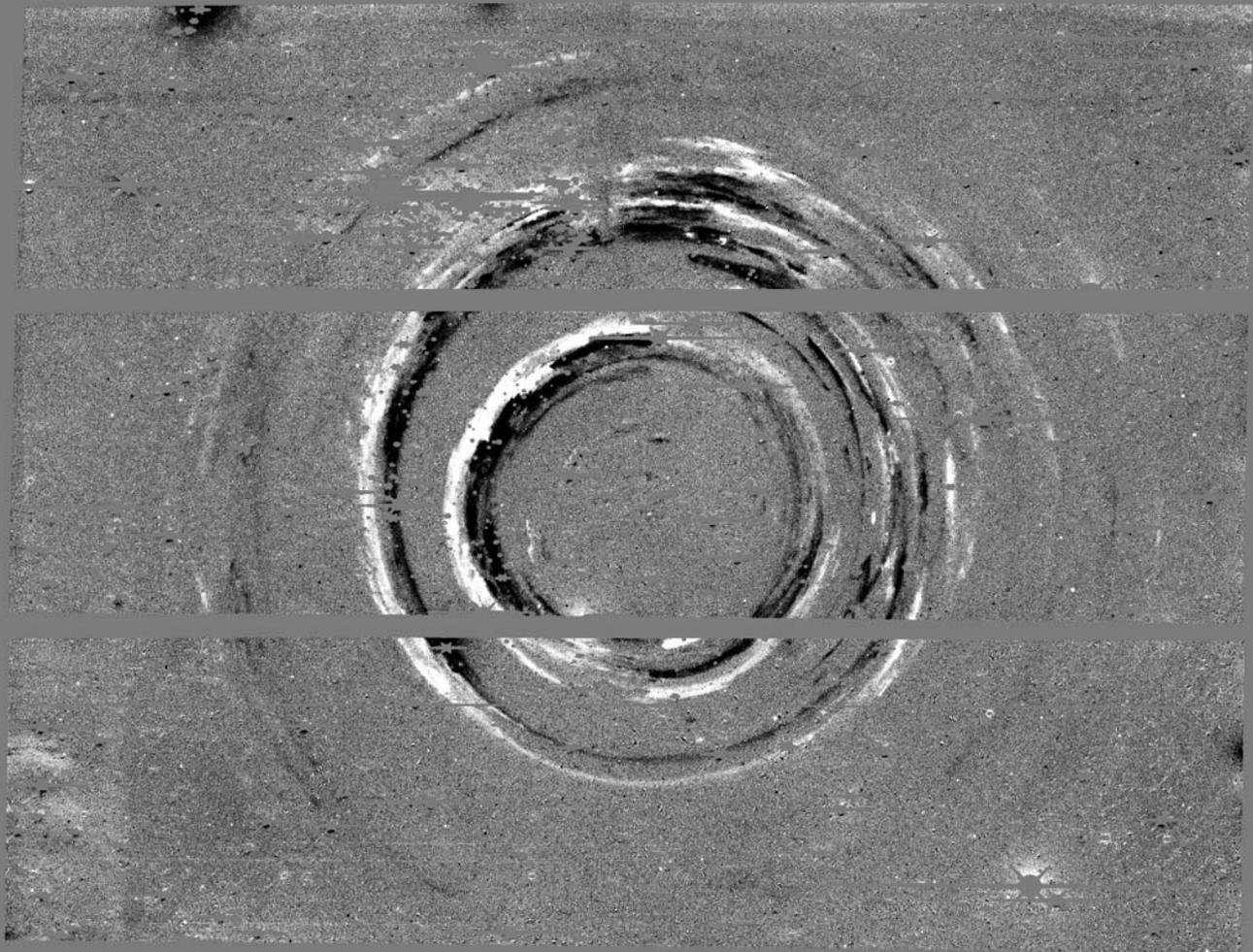

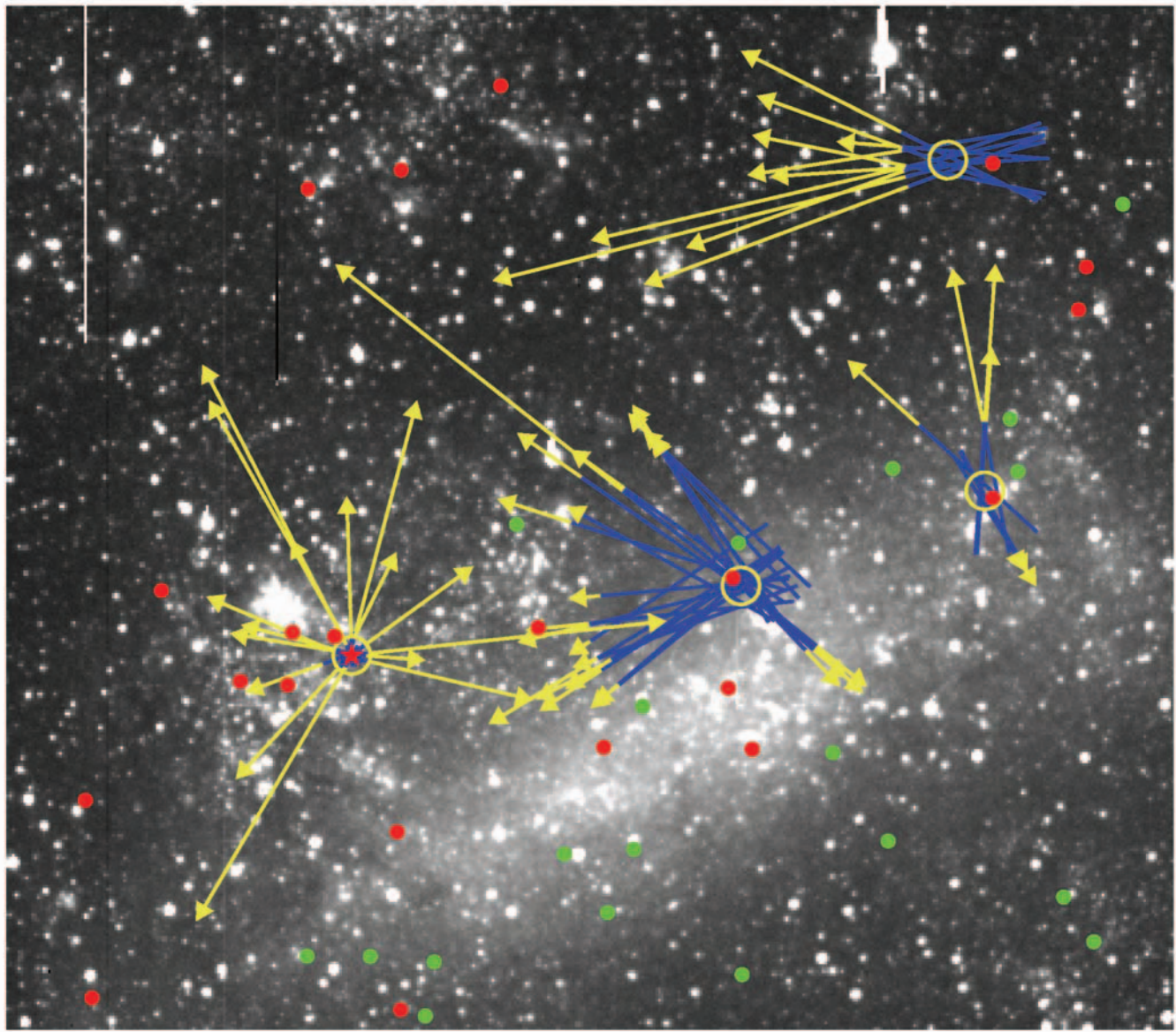

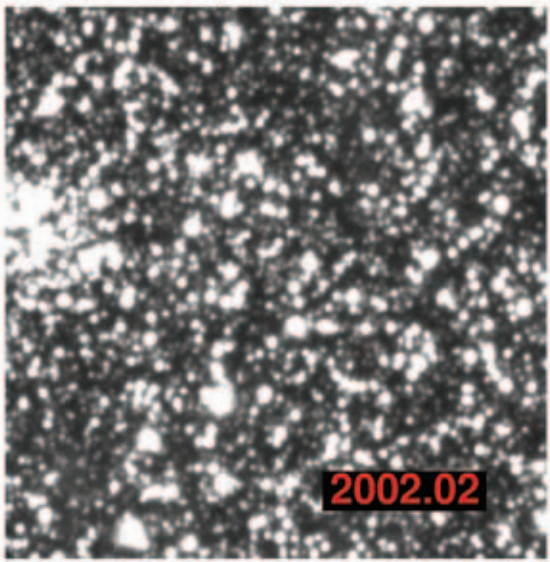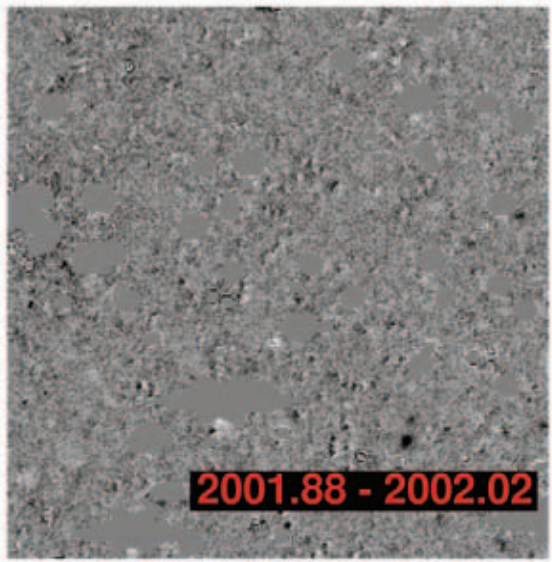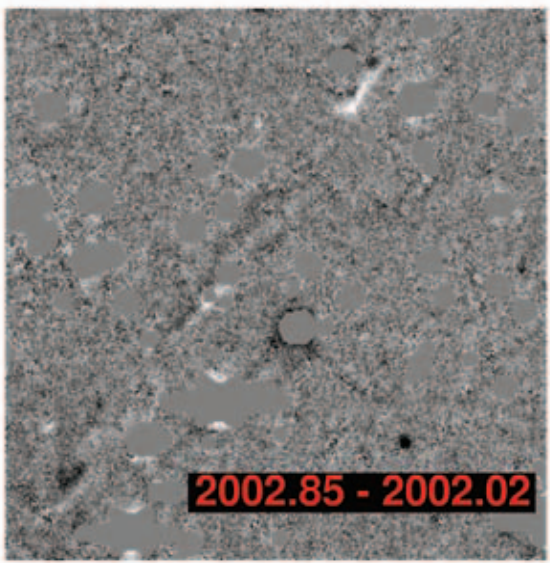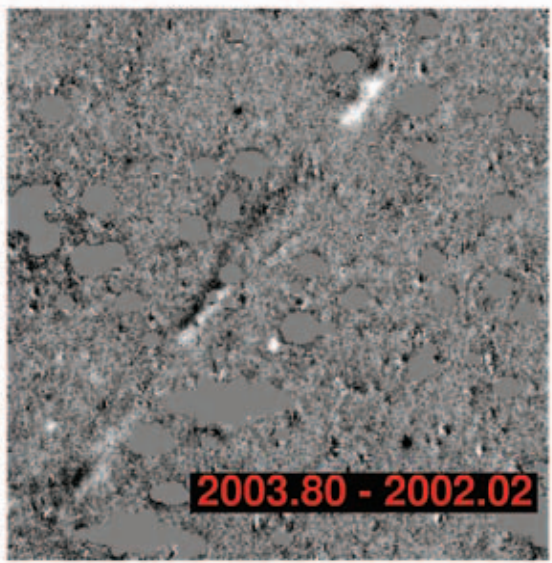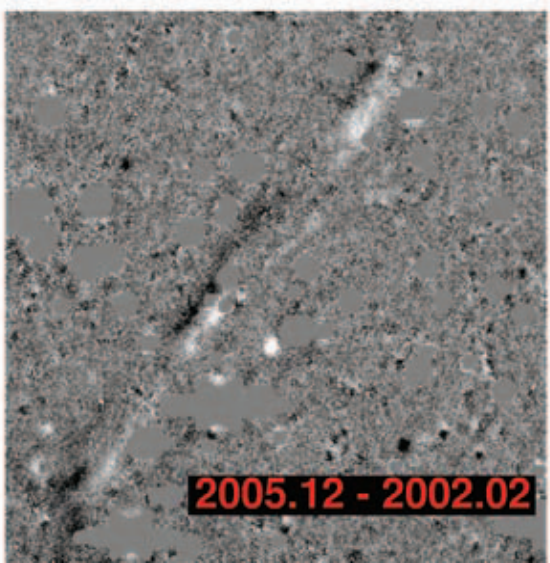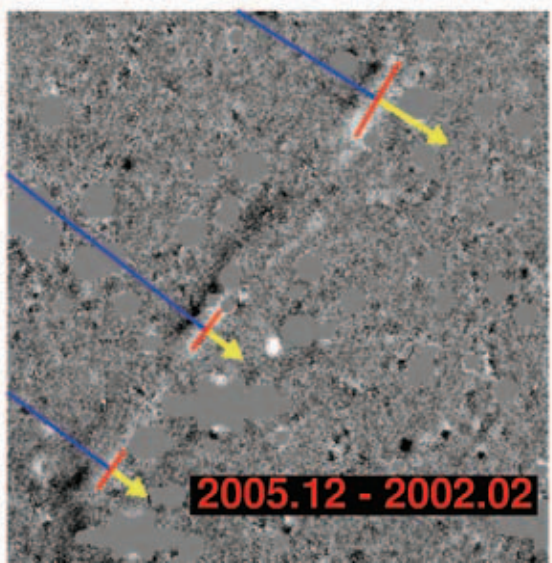